\documentclass[epj]{svjour}

\usepackage{amsmath}
\usepackage{latexsym}
\usepackage[dvips]{graphicx}
\usepackage[dvips]{epsfig}
\newcommand{\beq}{\begin{equation}}
\newcommand{\eeq}{\end{equation}}
\newcommand{\ds}{\displaystyle}

\newcommand{\dgr}{ {\,}^{\circ} \mbox{C}}
\newcommand{\un}[1]{\ensuremath{\unskip\,\mathrm{#1}}}

\begin{document}

\title{Electric Field Unbinding of Solid-Supported Lipid Multilayers}

\author{Doru Constantin\inst{1}\thanks{Present address~: Laboratoire de Physique de
l'\'{E}cole Normale Sup\'{e}rieure de Lyon, 46 all\'{e}e d'Italie, 69364 Lyon Cedex 07, France.} \and Christoph Ollinger\inst{1}%
\and Michael Vogel\inst{2} \and Tim Salditt\inst{1}}

\mail{D. C., \email{dcconsta@ens-lyon.fr} }

\institute{Institut f\"{u}r R\"{o}ntgenphysik, Geiststra{\ss}e 11,
37073 G\"{o}ttingen, Germany. \and Max-Planck-Institut f\"{u}r
Kolloid- und Grenzfl\"{a}chenforschung, D-14424 Potsdam, Germany.}

\date{Received: date / Revised version: date}
%
\abstract{ We studied by X-ray reflectivity the behaviour of fully hydrated solid-supported lipid
multilayers under the influence of a transverse electric field, under conditions routinely used in the
electroformation process. The kinetics of sample loss (unbinding) was measured as a function of the
amplitude and frequency of the applied field by monitoring the integrated intensity of the Bragg peaks. We
also performed a time-resolved analysis of the intensity of the first Bragg peak and characterized the
final state of the sample.
} 
\maketitle
\section{Introduction}
\label{sec:intro}

The influence of an external electric field upon lipid bilayers is of paramount importance, both from a
practical and a fundamental point of view. Cell electroporation \cite{neumann89} has been long employed for
introducing exogenous molecules into cells, while electroformation \cite{luisi00} is the best method to
date for obtaining giant unilamellar vesicles, which can be used as model systems for the cell membrane and
as transfection factors (see \cite{delattre93} and references therein). Although very important, these
phenomena are still in need of a clear-cut explanation. This lack can be understood in light of the fact
that both processes are instabilities involving large values of the electric field. As such, they lead to
large deformations, irreversibly changing the system structure~: creating pores in the membrane
(electroporation) or `peeling' the bilayers off the substrate (electroformation).

Here, we investigate the electroformation process
\cite{angelova92}, which involves applying an AC electric field
across a lipid film deposited on a substrate. The later stages of
the process, namely the formation of small vesicles and the
"ripening" leading to the appearance of giant vesicles have
already been studied in detail \cite{angelova92}. However, the
very first step, during which the lipid bilayers unbind from the
substrate, has not been quantified. In the present work, we focus
on characterizing this initial step, using the X-ray reflectivity
technique \cite{alsnielsen01}, which has the advantage of only
being sensitive to the bound membranes \cite{katsaras00}. Thus, we
can determine the quantity of lipid still left on the substrate at
a certain time (averaged over the fotprint of the beam) as well as
structural changes at the microscopic level (lamellar spacing,
fluctuation amplitude etc.) that are inaccessible by light
microscopy. It should be noted that the recent experimental work
of Burgess {\it et al.} \cite{burgess04} studies the influence of
the (static) surface charge on the structure of a unique supported
bilayer, using neutron reflectivity.

Several theories were put forward to explain different aspects of
the electric field effect upon the stability of lipid membranes
(flat \cite{thaokar02} or under the form of vesicles
\cite{isambert98,kumaran}, charged \cite{bensimon90,lau98,kim02}
or uncharged \cite{kummrow91}). To our knowledge, the only
theoretical paper relevant to our experimental configuration
(uncharged flat membrane under the influence of a transverse
electric field) is the one by Sens and Isambert \cite{sens02}.
They predict an undulation instability of the membrane, with a
typical wavelength in the submicron range.

\section{Experimental method}
\label{sec:exp}

\subsection{Sample preparation and environment}
\label{subsec:sample}

The field unbinding experiments were performed using the zwitterionic lipid
1,2-di\-myristoyl-sn-glycero-3-phos\-pho\-cho\-line (abbreviated in the following to DMPC), bought from
Avanti (Alabaster, AL, USA) and used without further purification. The lipid was dissolved in a 1:1
(vol/vol) mixture of chloroform and TFE (2- 2-2-trifluoroethanol) at concentrations between 5 and 20 mg/ml.
An amount of 0.1-0.2 ml of the solution was pipetted onto carefully cleaned silicon substrates of a size of
15 $\times$ 25 mm$^2$ cut from conductive (highly doped) commercial silicon wafers (Silchem Gmbh, Freiberg,
Germany). Depending on the desired film thickness, the samples were either left to dry slowly, in which
case all the lipid remains on the substrate, yielding stacks of 1000-3000 bilayers (depending on the
solution concentration and volume), or submitted to the spin-coating process \cite{mennicke02}. In this
case, immediately after deposition the substrate was accelerated to rotation (3000 rpm), using a
spin-coater. In this case, the number of bilayers $5 < N <25$ can be controlled. In both cases, the samples
were left to dry under a laminar flux hood at room temperature for a few hours and then exposed to high
vacuum at $40 \, ^{\circ}\mathrm{C}$ overnight to remove any remaining traces of solvent. They were finally
stored at $4 \, ^{\circ}\mathrm{C}$ until the measurement.

\begin{figure}[h!tbp]
\centerline{\epsfig{file=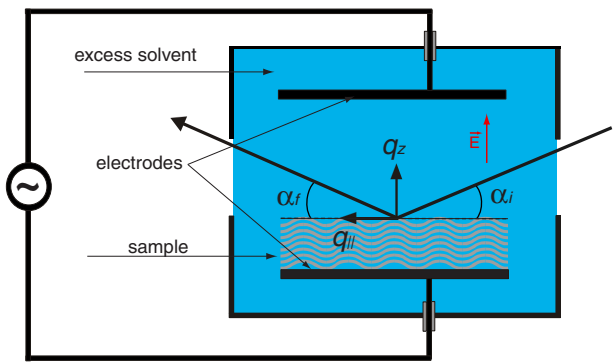,width=8cm}} \caption{Sketch of the sample chamber.} \label{Setup}
\end{figure}

The experiments were performed in a plexiglas chamber with kapton
windows (Fig. \ref{Setup}). The sample contacts and the counter
electrode are in stainless steel. The electric field is applied
using a function generator (Agilent Technologies). The
peak-to-peak amplitude is given in the text as Vpp. The cell was
mounted on a metal heating stage, temperature-con\-trolled by
water flow from a heating bath (Julabo Gmbh, Seelbach, Germany).
After mounting the sample, the hydrating solution was gently added
to avoid washing the lipid film off the substrate. We used
ultrapure water (Millipore, Bedford, MA).

\subsection{X-ray reflectivity}
\label{subsec:reflect}

The X-ray reflectivity measurements presented here were carried out at the bending magnet beamline D4 of
HASYLAB-DESY in Hamburg, Germany. At D4, a single-reflection Si(111) mono\-chro\-ma\-tor was used to select
a photon energy of 20 keV, after passing a Rh mirror to suppress higher harmonics. Primary beam intensity
is around $5 \, 10^8$ counts/s. The use of high energy X-rays has two reasons~: it reduces the beam damage
to the sample and minimizes absorbtion in the solvent (the transmission factor is about 0.25 for 2 cm of
water).

The chamber was mounted in horizontal scattering geometry, and the reflected beam was measured by a fast
scintillation counter (Cyberstar, Oxford), using computer-controlled aluminum absorbers which attenuate the
beam at small $q_z$ to prevent detector saturation. Incident and exit beams were defined by a system of
motorized slits. The data was corrected for decreasing electron ring current and for the diffuse
contribution (by subtraction of an offset scan). Finally, an illumination correction was performed.

\subsection{Time correlation function}
\label{subsec:corr}

For some samples we determined the autocorrelation function of the scattering signal. After aligning the
sample (usually with the first Bragg peak in specular condition), the signal from the scintillation
detector was fed into a multiple-tau correlator (ALV-5000/EPP, from ALV Gmbh, Langen, Germany) and the
corelation function \cite{berne00} $g(t)=\frac{\left \langle I(\tau)I(\tau+t)\right \rangle _{\tau}}{\left
\langle I(\tau)\right \rangle ^2}$ was obtained. $g(t)$ was averaged over 300 s. We performed these
measurements both on thick (spread) and thin (spin-coated) samples. No detectable modulation was present
for thick samples; however, for thin samples and at moderate frequencies, $g(t)$ exhibited an oscillation
at the frequency of the applied field. When this feature was observed, care was taken to rule out any
artefacts~: we checked that the modulation disappeared when unplugging the cable and when the sample
chamber was vigorously flushed with solvent to remove the sample from the substrate, maintaining the
experimental configuration and the applied field. This latter test conclusively shows that the oscillation
is indeed a feature of the lipid bilayers, and not of the substrate or mounting.

\section{Results}
\label{sec:results}

\subsection{Field amplitude}
\label{subsec:ampl}

\begin{figure*}[Htbp]
\centerline{\epsfig{file=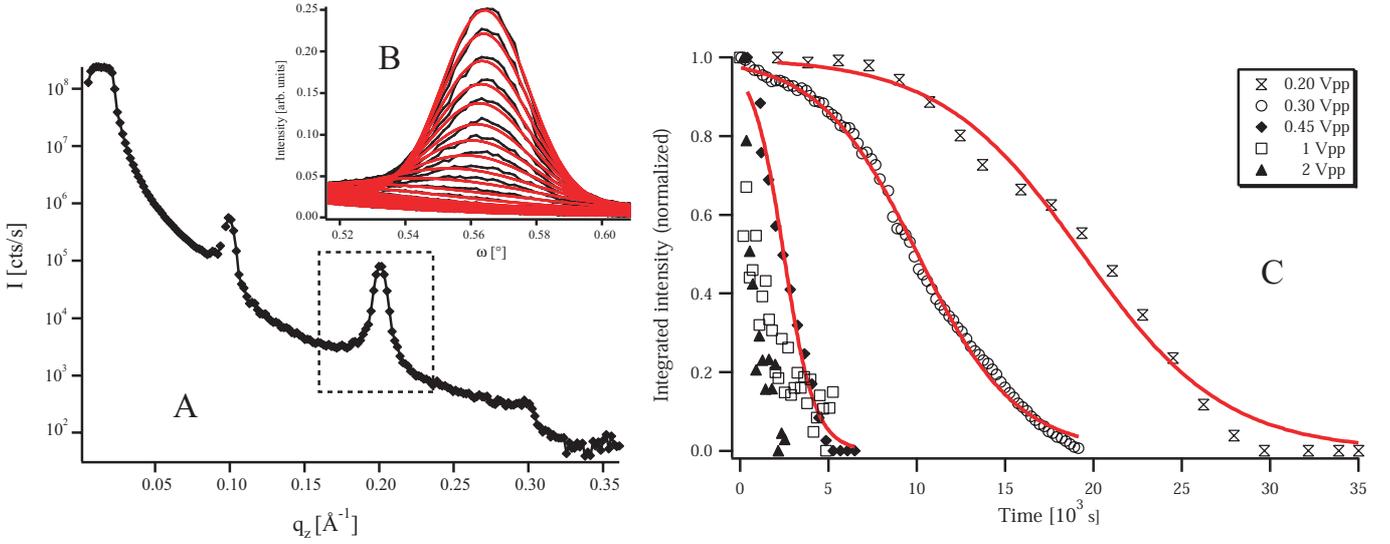, width=18cm}} \caption{A~: X-ray reflectivity curve of a thin DMPC
sample on Si substrate. B~: Zoom of the region around the second Bragg peak (dashed line in A), at
different times, under an applied electric field of 0.45 Vpp and 10 Hz. C~: Integrated intensity of the
second Bragg peak as a function of time, for different amplitudes of the electric field. For the three
lowest voltages, the fit with equation (\ref{eq:tanh}) is also shown (solid line).} \label{Time_evol}
\end{figure*}

We studied the unbinding kinetics as a function of the amplitude of the applied electric field. All the
measurements discussed in this subsection were performed at a fixed field frequency $\nu = 10 \un{Hz}$ and
at a temperature of $35 \dgr$. First, we present results obtained at low amplitudes, $U \leq 2 \un{Vpp}$,
on spin-coated samples. Figure \ref{Time_evol} shows a typical reflectivity spectrum before applying the
field (A). After the field is turned on, only the vicinity of a Bragg peak is scanned, as shown in the
inset (B). For all the curves, the measurements were performed on the second Bragg peak, where the
background (due to the interference with the substrate) is flatter and can be more easily subtracted. The
area under the Bragg peak is then integrated and corrected for the time variation of the primary beam.
Since the integrated intensity is roughly proportional to the sample volume (number of layers $\times$
illuminated area), we can thus monitor the kinetics of unbinding averaged over a few $\un{mm}^2$ of sample
area. The integrated intensity is shown in panel (C) as a function of time, for different values of the
applied field $U =$ 0.2, 0.3, 0.45, 1.0 and 2.0 Vpp. The longest evolution measured (at 0.2 Vpp) was
recorded over almost ten hours.

For the lowest applied fields, the intensity decay with time has a sigmoidal shape, characterized by a
"waiting time" and a "decay time"; we quantified this tendency by fitting the curves with a hyperbolic
tangent~:

\beq I(t)=\frac{1}{2} \left [ 1 - \tanh \left ( \frac{t-t_0}{\Delta t} \right ) \right ] \label{eq:tanh}
\eeq

yielding the following values~:

\begin{table}[htbp]
\begin{center}
\begin{tabular}{c|cccc}
         $U$ [Vpp] & $t_0$ [s] & $\Delta t$ [s] & Sample & \\ \hline
         0.2 & 20000 $\pm$ 300 & 8000 $\pm$ 500 & thin & Fig. \ref{Time_evol} C\\
         0.3 & 10000 $\pm$ 40 & 5500 $\pm$ 60 & -"- & -"-\\
         0.45 & 2500 $\pm$ 70 & 1800 $\pm$ 120 & -"- & -"-\\
         1.0 & 0 [fixed] & 1500 $\pm$ 250 & -"- & -"-\\
         2.0 & 0 [fixed] & 800 $\pm$ 100 & -"- & -"-\\
         5.0 & 0 [fixed] & 1300 $\pm$ 100 & -"- & not shown\\
         1.0 & 3200 $\pm$ 80 & 1500 $\pm$ 120 & thick & not shown\\
\end{tabular}
\end{center}
\caption{Parameters of the intensity decay as a function of the amplitude of the field (at a fixed
frequency $\nu = 10 \un{Hz}$) fitted with equation \ref{eq:tanh}.}
\end{table}

At higher field amplitudes (above 1Vpp), the "waiting time" vanishes, so that we set $t_0=0$. The remaining
decay could be equally well described by an exponential function.

We also studied thick samples under the same experimental
conditions. For comparison, the evolution of the scattered
intensity was fitted using the same model as for the thin samples
(equation \ref{eq:tanh}) and the parameters are given in Table 1.

The main interest of using thick samples is, however, the
possibility of studying the diffuse scattering signal and of
obtaining information about the bilayer position fluctuations.
Extensive measurements were performed, both at D4 (HASYLAB) and at
the ID1 beamline at the ESRF (Grenoble, France); no difference was
detected in the diffuse spectrum between the situation with or
without electric field (data not shown). We therefore conclude
that the bilayers within the sample are not significantly affected
by the presence of the field (no enhancement in the fluctuation
amplitude).

\subsection{Comparison with thermal effects}
\label{subsec:thermal}

\begin{figure}[h!tbp]
\centerline{\epsfig{file=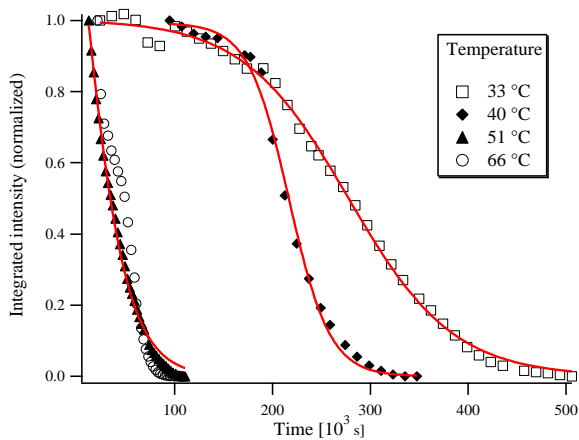,width=8cm}} \caption{Integrated intensity of the first Bragg sheet
(diffuse signal) of an OPPC sample as a function of time, for different temperatures. For the three lowest
temperatures, the fit with equation (\ref{eq:tanh}) is also shown (solid line).} \label{thermal}
\end{figure}

The dependence of the unbinding rate on the amplitude of the electric field is strikingly similar to the
kinetics recorded as a function of the temperature \cite{vogel00}. The same pattern of a plateau followed
by a decay (at lower temperature) and the disappearance of the plateau (at higher temperatures) emerge,
albeit on a completely different time scale (see Figure \ref{thermal}).

The lipid used was 1-oleoyl-2-palmitoyl-sn-glycero-3-phos\-pho\-cho\-line (OPPC), bought from Avanti
Lipids, Alabama. The samples were prepared as for the electric field studies (by spreading), except for
using isopropanol as a solvent and for applying a hydrophilizing treatment to the substrates, by washing
them in a saturated KOH solution in ethanol for about a minute. Subsequently, they were rinsed several
times with ultra-pure water.

The data has been taken in a temperature-controlled steel chamber sample immersed in excess ultra-pure
water (the beam path in water is 15 mm), on an in-house setup using a sealed tube generator (Mo K$_\alpha$
radiation, $E=17.4 \un{keV}$) and a two-circle goniometer (Siemens D500). The integrated diffuse scattering
of the first Bragg sheet was used as a measure of the total scattering volume.

No high-temperature measurements were performed for DMPC, so that a direct comparison with the results of
subsection \ref{subsec:ampl} is not possible. However, several measurements were made using one DMPC
sample, at temperatures between 22 and $30 \dgr$, over a period of four weeks, without any apparent sample
loss.

\subsection{Field frequency}
\label{subsec:freq}

The temperature was set at $30 \dgr$ for these measurements.

\begin{figure}[h!tbp]
\centerline{\epsfig{file=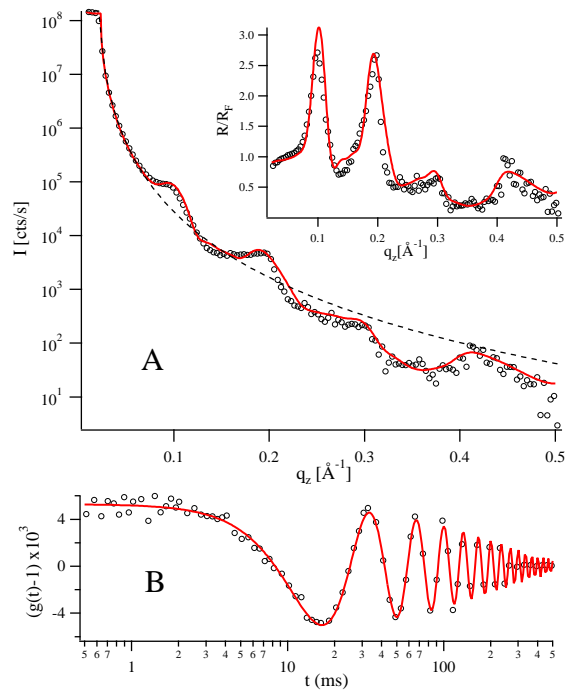,width=8cm}}
\caption{A~: Reflectivity of the spin-coated sample discussed in
the text (open symbols) adjusted with the model described in
\cite{salditt02,mennicke04} (red solid curve) and the associated
Fresnel reflectivity (dashed line). The inset shows the same data
scaled by the Fresnel reflectivity. B~: Correlation function
measured on the first Bragg peak ($q_z=0.101 \mbox{\AA}^{-1}$)
under an applied electric field with $\nu=30 \mbox{Hz}$ (open
symbols) and fit with a damped cosine (red solid curve). See text
for details.} \label{refl_corr_thin}
\end{figure}

A representative example is shown in Figure \ref{refl_corr_thin};
graph A shows the specular scan (corrected for the offset
contribution) and the adjustment with the model given in
\cite{salditt02,mennicke04} (red solid curve), as well as the
associated Fresnel reflectivity (dashed line). The inset shows the
same data divided by the Fresnel reflectivity. From the model, we
infer that the sample consists of four layers, with approximate
coverage ratios (starting at the substrate)~: 1.0, 0.8, 0.45, 0.15
(other sample parameters, such as the amplitudes of the different
Fourier components of the electron density profile and the
Caill\'{e} parameter $\ds \eta = \frac{\pi}{2} \frac{k_B T}{B
\lambda d^2}$ quantifying the amplitude of the thermal
fluctuations were taken from previous measurements under the same
conditions \cite{mennicke04}). Graph B shows a time correlation
function measured on top of the first Bragg peak ($q_z=0.101
\mbox{\AA}^{-1}$, corresponding to a repeat spacing $d=62.0
\mbox{\AA}$), under an applied electric field with an amplitude of
5 Vpp and a frequency $\nu = 30 \un{Hz}$. The correlation function
is well described by a decaying oscillation at the applied
frequency~:

\beq g(t)-1=g_0 \cos (2 \pi \nu t) \exp(-t/\tau) \label{eq:corr} \eeq

\noindent where the amplitude $g_0 \simeq 5\, 10^{-3}$ and the
decay time $\tau = 0.23 \un{s}$. Similar measurements were
performed at slightly different frequencies; the behaviour of
$g(t)$ follows the applied frequency, ruling out the presence of
an external oscillator. As discussed above (subsection
\ref{subsec:corr}), we took additional step to confirm that the
signal is indeed a genuine contribution of the sample.

We recorded the correlation function on top of the first Bragg
peak for increasing frequencies and noticed that, above a
threshold of the order of $\nu_{\mathrm max}= 100 \un{Hz}$,
estimated from measurements performed on several samples, $g(t)$
no longer exhibits oscillations at the applied frequency.

In the range where the correlation function can be properly
measured (about 10 to 50 Hz), the recorded frequency $\nu$ is very
precisely that of the applied electric field. The decay time
exhibits an intriguing dependence on $\nu$~: $\tau \sim 7 / \nu$,
but more systematic measurements would be needed to confirm
whether this is an intrinsic feature of the sample or an artifact
of the setup.

\begin{figure}[h!tbp]
\centerline{\epsfig{file=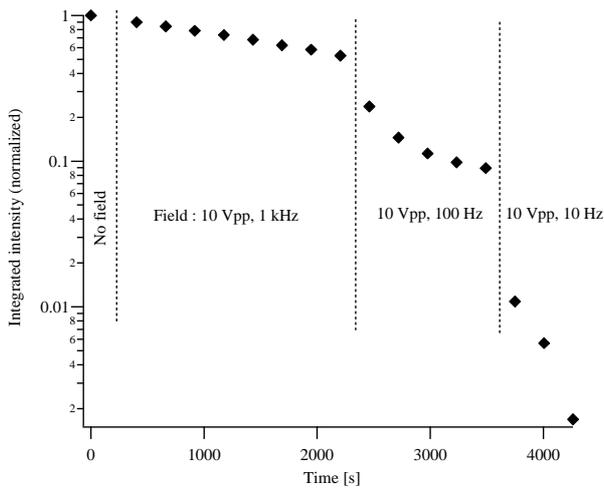,width=8cm}} \caption{Unbinding kinetics for a thick (spread)
sample, as a function of the frequency of the applied electric field. Note the logarithmic intensity
scale.} \label{freq_effect}
\end{figure}

The frequency of the applied field also has a strong effect on the
unbinding kinetics, as can be seen in Figure \ref{freq_effect}.
The measurement was measured on a thick (about 3000 bilayers)
sample. We started by applying a 10 Vpp field at 1 kHz and then
decreased the frequency at 100 Hz and finally at 10 Hz. The change
in the decay rate of the peak is very strong. In addition, an
abrupt decrease in intensity was observed when switching from 100
to 10 Hz; this effect is systematically observed in all samples
studied at high field amplitude (data not shown), and could be
related to the unbinding of weakly adhering surface patches.

\subsection{Final state}
\label{subsec:final}

\begin{figure}[h!tbp]
\centerline{\epsfig{file=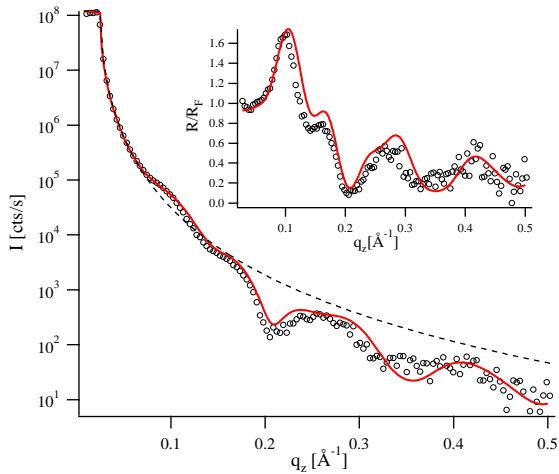,width=8cm}} \caption{Reflectivity of a spin-coated sample in the
final state, after applying a strong electric field (open symbols) adjusted with the model described in
\cite{mennicke04} (red solid curve) and the associated Fresnel reflectivity (dashed line). The inset shows
the same data scaled by the Fresnel reflectivity.} \label{final_refl}
\end{figure}

An interesting point to address is also the final state, defined
as the point at which applying a reasonably strong electric field
no longer has any observable effect on the sample. In most of the
measurements performed on thin (spin-coated) samples, it appears
that only one or two bilayers remain on the substrate. An example
is shown in Figure \ref{final_refl}, where a sample prepared
according to the protocol in \cite{mennicke02} using a DMPC
concentration of 2 mg/ml was exposed to an AC field with an
amplitude of 10 Vpp at a frequency of 10 Hz, and afterwards at 5
Hz for more than 10 minutes in each case; the final reflectivity
(open symbols) can be described by assuming that only two bilayers
are left on the substrate, with coverage ratios of 0.8 and 0.3,
starting at the substrate. This spectrum being a representative
one for spin-coated samples, we can conclude that the bottom layer
(closest to the substrate) is largely unaffected by the electric
field.

\section{Discussion and Conclusion}
\label{sec:conc}

At the beginning of this study, two kinds of results were
expected~: long-time effects (unbinding kinetics under electric
field) and changes in the short-time behaviour (the fluctuation
spectrum) of the lamellar phase.

{\bf Unbinding kinetics}~: The unbinding of lipid bilayers from a
solid substrate represents the first step of the electroformation
process. We determined the kinetics of unbinding under field for
various applied voltages, both for thin (tens of bilayers) and
thick (thousands of bilayers) samples. A two-step process was
observed at low voltage, consisting in a very slow decay
("incubation" plateau) followed by a fast sample loss. At higher
applied voltages, the first step disappears. Surprising
similarities with the previously characterized thermal unbinding
kinetics were thus revealed. The frequency of the applied field is
also important: the unbinding no longer occurs when the frequency
exceeds a few hundred Hz.

The field-induced instability discussed here and the
disintegration pathway of the lamellar stack has been corroborated
in many sample series of DMPC, at varied field amplitudes, and
frequencies, both for thin and thick stacks \cite{ollinger03}.
Moreover, it was found that the presence of an osmotic stressor
(polyethylene glycol, MW 20000) in solution does not prevent the
instability, but does slow down the decay rate \cite{ollinger03}.
However, the results of measurements on thick stacks and/or in the
presence of a osmotic stress being rather noisy, we do not discuss
them here.

{\bf Fluctuations}~: As we know that the applied field induces the
unbinding, it must necessarily have an effect at the microscopic
scale (i.e. that of the individual membrane). This effect could be
at the level of the structural parameters (changes in the bilayer
and/or water layer thickness) but could also involve enhanced
fluctuations of the bilayers, as invoked in the case of thermal
unbinding. In addition, recent theoretical work \cite{sens02} has
suggested that a transverse electric field could induce an
undulation instability of a (low conductivity) membrane.

We could detect no change in the structural parameters or in the
diffuse scattering for thick samples; in thin samples, on the
other hand, the intensity of the Bragg reflection exhibits a
modulation at the frequency of the applied field. This leads us to
the tentative conclusion that the electric field only affects a
few layers at the top of the stack and that, in thick samples,
their contribution to the scattered signal is swamped by that of
the bulk. It appears that we are dealing with a surface
phenomenon, which is reasonable in light of the fact that the
electroformation phenomenon yields unilamellar vesicles; at some
point, the bilayers must peel off one by one.

This conclusion is not in contradiction with the theoretical model
of Sens and Isambert \cite{sens02}, since they also observe a
strong change in dynamics in the presence of fixed boundaries. It
would be reasonable to assume that motion of the membranes within
the stack is hindered by their neighbours (and the boundary
condition at the sustrate), while the top ones are relatively free
to fluctuate. A more quantitative comparison cannot be made, since
we cannot determine accurately the number of affected bilayers at
the surface, the fluctuation amplitude and length scales; it
should also be noted that membrane integrity is a key ingredient
of their model (the electric current flows across the membrane,
and the transverse electric field is amplified in proportion to
the resistivity of the membrane).

This condition is not fulfilled in our samples, where defects
(dewetted patches) are known to appear \cite{perrinogallice02}.
The electric current is more likely to flow through the defects,
rather than across the bilayers; this is also the case of
electroformation procedures, where the deposited lipid film need
not cover the entire surface of the electrode \cite{angelova92}.

We conclude that the presence of the electrodynamical instability
studied by Sens and Isambert \cite{sens02} is neither confirmed
nor can it be refuted by our experiments; a more careful setup,
involving {\it e. g.} a freely suspended lipid film, would be
needed.

In these conditions, an alternative explanation could be related
to the presence of electro-osmotic flow in the solution, but more
comprehensive tests must be performed in order to test this
hypothesis.



\begin{acknowledgement}
D. C. has been supported by a Marie Curie Fellowship of the
European Community programme \textit{Improving the Human Research
Potential} under contract number HPMF-CT-2002-01903. Chenghao Li
is acknowledged for helping with the D4 experiments, and Manouk
Abkarian for assistance with the electroformation protocol.
\end{acknowledgement}

\end{document}